\documentclass[11pt]{article}

\usepackage[final]{acl}

\usepackage{times}
\usepackage{latexsym}

\usepackage[T1]{fontenc}
\usepackage{tcolorbox}

\usepackage[utf8]{inputenc}

\usepackage{microtype}

\usepackage{inconsolata}

\usepackage{graphicx}
\usepackage{booktabs}

\title{Augmenting Researchy Questions with Sub-question Judgments}

\author{
 \textbf{Jia-Huei Ju\textsuperscript{1}},
 \textbf{Eugene Yang\textsuperscript{2}},
 \textbf{Trevor Adriaanse\textsuperscript{3}},
 \textbf{Andrew Yates\textsuperscript{2}},
\\
 \textsuperscript{1}University of Amsterdam, \\
 \textsuperscript{2}Human Language Technology Center of Excellence, Johns Hopkins University, \\
 \textsuperscript{3}Johns Hopkins University,
\\
 \small{
   \textbf{Correspondence:} \href{mailto:j.ju@uva.nl}{j.ju@uva.nl}
 }
}

\begin{document}
\maketitle
\begin{abstract}
The Researchy Questions dataset provides about 100k question queries with complex information needs that require retrieving information about several aspects of a topic.
Each query in ResearchyQuestions is associated with sub-questions that were produced by prompting GPT-4.
While ResearchyQuestions contains labels indicating what documents were clicked after issuing the query, there are no associations in the dataset between sub-questions and relevant documents.
In this work, we augment the Researchy Questions dataset with LLM-judged labels for each sub-question using a Llama3.3 70B model.
We intend these sub-question labels to serve as a resource for training retrieval models that better support complex information needs. 
\end{abstract}

\section{Introduction}

Researchy Questions~\cite{rosset2024researchy} is a collection of queries taken from the Bing query log, along with the URLs that the user clicked, which are also present in the ClueWeb22 dataset~\citep{overwijk2022clueweb22}. 
Such questions are complex and usually require searching with multiple queries in a single session to fully answer. 
In the original dataset, the authors provide the decomposition of each question into several sub-questions by GPT4. 
However, there are no relevance judgments linked to each decomposed sub-questions, which limit the usability of the collection as a sub-question training resource. 

To fill the gap, we employ an automatic judgment model to generate the relevance labels for the training sub-questions. 
We adapt a graded judgment prompt from CRUX~\cite{ju2025controlled} to capture different granularities of the relevance. 
Compared to typical retrieval training collections such as MS MARCO~\cite{bajaj2016ms}, sub-questions provided by Researchy Qeustions are less factual and require better semantic understanding to the questions and the documents. 
Such an addition provides a concrete usage for the original Researchy Question collection for training factoid question retrieval model.

\begin{figure}
\begin{tcolorbox}[title=Qwen3-Reranking Instruction]
\textbf{System prompt}: 
Judge whether the Document meets the requirements based on the Query and the Instruct provided. Note that the answer can only be "yes" or "no". \\
\textbf{Prompt}: 
\textless Instruct\textgreater: Given the list of questions as query, retrieve relevant passages that answer the questions.
\textless Query\textgreater: \{{\rm join}(subquestions)\}
\textless Document\textgreater: \{doc\}"
\end{tcolorbox}
\caption{The prompt used to rerank documents with the pointwise Qwen3 0.6B reranking model.} 
\label{fig:rerank}
\end{figure}

\begin{figure*}[tb]
\begin{tcolorbox}[title=Rubric-based Answerability Judgment]
Instruction: Determine whether the question can be answered based on the provided context? Rate the context with on a scale from 0 to 5 according to the guideline below. Do not write anything except the rating. \\\\
Guideline: \\
5: The context is highly relevant, complete, and accurate. \\
4: The context is mostly relevant and complete but may have minor gaps or inaccuracies.\\
3: The context is partially relevant and complete, with noticeable gaps or inaccuracies.\\
2: The context has limited relevance and completeness, with significant gaps or inaccuracies.\\
1: The context is minimally relevant or complete, with substantial shortcomings.\\
0: The context is not relevant or complete at all. \\\\
Question: \{$q$\} \\ Context: \{$c$\} \\ Rating:
\end{tcolorbox}
\caption{The prompt used to assess the relevance of a document (context) \textit{c} to a sub-question \textit{q} on a scale from 0 to 5, following prior work obtaining sub-question judgments~\citep{ju2025controlled}.}
\label{fig:judge}
\end{figure*}

\section{Candidate Documents}

To avoid indexing the entire ClueWeb22~\citep{overwijk2022clueweb22}, which contains more than 10B documents, we only consider documents in the English subset of ClueWeb22 Set B, which contains 48\% of the clicked documents in Researchy Questions. 
For efficiency, we retrieved documents from ClubWeb22 Set B using BM25. 
To avoid overly biasing results toward lexical matches, we reranked the top 100 documents with the Qwen3 0.6B Reranker~\cite{qwen3embedding} using a slightly modified instruction documented in Figure~\ref{fig:rerank}.

For selecting the candidate documents for each sub-question of the Researchy Questions Train split, we collect the top 20 documents from the initial BM25 retrieval results and the top 20 documents from the reranking output to balance the lexical and semantic matching in the candidate set. 
We additionally include the clicked documents (439,161 in total) of the original question supplied by the Researchy Question collection to ensure they are also processed by our downstream pipeline. 
These documents are then passed to an automatic judgment model for acquiring relevance labels, which we describe in the next section.

\section{Automatic Judgments}

After identifying candidate documents, we used the Llama3.3 70B model to produce relevance judgments.
Each document was automatically judged for each sub-question derived from the top-level query. 
Judgments were on a 0-5 scale using the prompts from from~\citet{ju2025controlled}, which are intended to assess the answerability~\citep{dietz2024workbench,farzi2024exam} of a question given a document.
The prompt is shown in Figure~\ref{fig:judge}.

\section{Dataset Statistics}

For all 1,288,976 sub-questions in the Researchy Question training set, we acquire 24,316,320 predictions, which on average have 18.87 judgments per sub-question. 
Table~\ref{tab:stats} summarizes the distribution of the documents predicted in each relevance level.

\begin{table}[]
    \centering
    \begin{tabular}{l|rc}
    \toprule
    Level & Count     & Per Sub-question \\
    \midrule
    5     & 1,907,722 & 1.48  \\
    4     & 5,444,774 & 4.22  \\
    3     & 2,909,134 & 2.26  \\
    2     & 7,920,642 & 6.14  \\
    1     & 2,651,382 & 2.06  \\
    0     & 3,482,653 & 2.70  \\
    \midrule
    Other &        13 &   --  \\
    \bottomrule
    \end{tabular}
    \caption{Statistics of each predicted relevance level. }
    \label{tab:stats}
\end{table}

\section{Conclusion}

In this work, we augment the Researchy Question dataset with automatic labels using CRUX. 
This resource will enable modeling exploration for factoid retrieval and question answering. 
We provide these generated labels as a community resource for encouraging research in these directions. 

\bibliography{custom}

\begin{thebibliography}{7}
\providecommand{\natexlab}[1]{#1}

\bibitem[{Bajaj et~al.(2016)Bajaj, Campos, Craswell, Deng, Gao, Liu, Majumder,
  McNamara, Mitra, Nguyen et~al.}]{bajaj2016ms}
Payal Bajaj, Daniel Campos, Nick Craswell, Li~Deng, Jianfeng Gao, Xiaodong Liu,
  Rangan Majumder, Andrew McNamara, Bhaskar Mitra, Tri Nguyen, and 1 others.
  2016.
\newblock {MS} {MARCO}: A human generated machine reading comprehension
  dataset.
\newblock \emph{arXiv preprint arXiv:1611.09268}.

\bibitem[{Dietz(2024)}]{dietz2024workbench}
Laura Dietz. 2024.
\newblock A workbench for autograding retrieve/generate systems.
\newblock In \emph{Proceedings of the 47th International ACM SIGIR Conference
  on Research and Development in Information Retrieval}, pages 1963--1972.

\bibitem[{Farzi and Dietz(2024)}]{farzi2024exam}
Naghmeh Farzi and Laura Dietz. 2024.
\newblock An exam-based evaluation approach beyond traditional relevance
  judgments.
\newblock \emph{arXiv preprint arXiv:2402.00309}.

\bibitem[{Ju et~al.(2025)Ju, Verberne, de~Rijke, and Yates}]{ju2025controlled}
Jia-Huei Ju, Suzan Verberne, Maarten de~Rijke, and Andrew Yates. 2025.
\newblock {Controlled Retrieval-augmented Context Evaluation for Long-form
  RAG}.
\newblock \emph{arXiv preprint arXiv:2506.20051}.

\bibitem[{Overwijk et~al.(2022)Overwijk, Xiong, Liu, VandenBerg, and
  Callan}]{overwijk2022clueweb22}
Arnold Overwijk, Chenyan Xiong, Xiao Liu, Cameron VandenBerg, and Jamie Callan.
  2022.
\newblock Clueweb22: 10 billion web documents with visual and semantic
  information.
\newblock \emph{arXiv preprint arXiv:2211.15848}.

\bibitem[{Rosset et~al.(2024)Rosset, Chung, Qin, Chau, Feng, Awadallah,
  Neville, and Rao}]{rosset2024researchy}
Corby Rosset, Ho-Lam Chung, Guanghui Qin, Ethan~C Chau, Zhuo Feng, Ahmed
  Awadallah, Jennifer Neville, and Nikhil Rao. 2024.
\newblock {Researchy Questions: A dataset of multi-perspective, decompositional
  questions for llm web agents}.
\newblock \emph{arXiv preprint arXiv:2402.17896}.

\bibitem[{Zhang et~al.(2025)Zhang, Li, Long, Zhang, Lin, Yang, Xie, Yang, Liu,
  Lin, Huang, and Zhou}]{qwen3embedding}
Yanzhao Zhang, Mingxin Li, Dingkun Long, Xin Zhang, Huan Lin, Baosong Yang,
  Pengjun Xie, An~Yang, Dayiheng Liu, Junyang Lin, Fei Huang, and Jingren Zhou.
  2025.
\newblock Qwen3 embedding: Advancing text embedding and reranking through
  foundation models.
\newblock \emph{arXiv preprint arXiv:2506.05176}.

\end{thebibliography}

\end{document}